# Accurate FTIR determination of boron concentration in CVD homoepitaxial diamond layers

*Mikhail Panov*[1], *Vasily Zubkov*[1], *Anna Solomnikova*[1], *Igor Klepikov*[2,3]

[1] St. Petersburg Electrotechnical University "LETI", 197022, St. Petersburg, Russia

[2] St. Petersburg State University, 199034, St. Petersburg, Russia

[3] LLC NPK Almaz, Saint-Petersburg, 197706, Russia

**Abstract**

The intensive development of technology for fabrication semiconducting CVD diamond layers poses an important task of developing a precise and non-destructive method for estimation the boron content in thin epitaxial layers. For bulk and uniformly doped diamond samples, the infrared optical spectroscopy successfully performs such a role. Here we propose a correct method to determine the boron concentration in CVD homoepitaxial diamond layers from FTIR spectra. The method is the natural advancement of the existing technique for bulk samples. The feature of the novel technique is the accurate accounting of passing radiation through a multilayered structure with different thicknesses of absorbing media for special absorbing mechanisms. For this situation, an expression for the effective optical density is obtained. We have demonstrated the benefit of the method for a set of samples with CVD homoepitaxial layers grown on various HPHT substrates with and without nitrogen impurity. The measured FTIR spectra were subdivided into relevant sections responsible for the specific absorption mechanisms, and the correct amplitudes of the boron absorption peaks were derived. The data obtained from FTIR spectra is thoroughly compared to the charge carrier concentration derived from electrical capacitance-voltage measurements.

## Introduction

Measurements of impurity concentration in diamond crystals by optical methods have a long history [1–5]. Due to the extremely wide bandgap ($E_g$ = 5.45 eV at room temperature [6,7]), undoped diamond is a dielectric and is transparent in the entire visible range of the electromagnetic spectrum. Since this material possesses clearly expressed dielectric properties and is being actively used in gemology and jewelry, the stones classification is conventionally carried out according to the impurity concentration derived from nondestructive optical methods.

For almost a hundred years diamonds are being conditionally classified into types I and II according to their transparency to UV radiation. Type I diamonds do not transmit radiation with a wavelength λ less than 300 nm, while type II diamonds are transparent to UV radiation with λ up to 230 nm. It has long been discovered that diamonds, containing boron impurity (type IIb), possess semiconductor properties [8,9]. The blue color of boron-doped samples is explained by intense absorption of the red and infrared parts of light spectrum falling on the crystal [10]. Unfortunately, the blue crystals containing boron are very rare among natural diamonds (less than 0.1% [11,12]). King et al. tried to establish the dependence between the intensity of the blue color of a sample and its characteristics (intensity and duration of UV phosphorescence, electrical conductivity, etc.) [13]. The authors [14,15] studied the temperature dependence of electrical resistance of *p*-type synthetic diamonds at different boron content in the growth mixture. In 1971, A. Collins (King's College, London) was the first to propose an optical technique for reliable determination of the concentration of compensated boron impurity in semiconductor diamond [2], which corresponded to the color of the crystal and correlated with the results of other measurements.

At present, the updated and expanded Collins calibration coefficients for two boron IR absorption bands are used to estimate the concentration of partially compensated boron (understood as the difference of the acceptor and donor concentrations $N_A$–$N_D$) in *bulk* samples with uniform impurity distribution [11]:

$$[N_A–N_D](\text{ppm}) = 0.0350\times H_{2800},$$
$$[N_A–N_D](\text{ppm}) = 0.105\times H_{2458}, \tag{1}$$

where $H_{2800}$, $H_{2458}$ are the amplitudes of the corresponding peaks in the absorption spectrum calculated above a baseline. As it is known, for boron-doped diamond 1 ppm (parts per million) of impurity approximately corresponds to its concentration $[N_A–N_D] = 1.67\cdot10^{17}$ cm$^{-3}$. The value of the absorption coefficient α is determined from the experimentally measured relative transmittance $T$ according to

$$\alpha = \frac{1}{d}\ln\frac{(1-R)^2}{T}. \tag{2}$$

Here $d$ is the sample thickness, and $R$ is the sample reflectance.

Authors [11,16] suggested in 2019 the measurements of boron concentration distribution in plates of bulk diamond via the automated impurity mapping based on the above-cited calibration formulas (1) with lateral resolution of 100 μm.

Importantly, the results derived from optical transmittance measurements are fundamentally not absolute, since the analyzed radiation passes from the light source to the photodetector through a monochromator, condenser and other optical systems. Each of these elements has its own spectral transfer characteristic. Therefore, the results obtained should be calibrated to direct concentration measurements of absorbing impurity. The Hall effect measurements are primarily preferred for this purpose.

On the other hand, strictly speaking, Hall measurements determine the concentration of free charge carriers, but not impurities. So, Hall calibration for diamond (in contrast to that for traditional semiconductors), requires measurements at very high temperatures (up to 1250 K) to achieve the complete ionization of charge carriers (holes), as the degree of boron ionization is less than 1% at room temperature due to its high activation energy (370 meV) [2].

The intensive development of the technology for depositing CVD diamond epitaxial layers poses an urgent task of expanding the method of estimation the boron concentration by non-destructive IR spectroscopy to *epitaxial layers*. The pres-ence of two substantially dissimilar parts of the plate makes the Collins-Howell technique unsuitable.

**Problem statement. Theory**

Consider single-crystal diamond sample consisting of a boron-doped CVD epitaxial layer grown on an HPHT substrate, Fig. 1. When analyzing the impurity concentration using the common FTIR method, the probing radiation (with intensity $I_0$) passes through a two-layer structure, so it is necessary to separately take into account the absorption in the thick substrate and in the thin epitaxial layer. In this case, the standard formula (2), describing the relationship between the transmission coefficient $T$ and the boron absorption coefficient α, becomes more complicated.

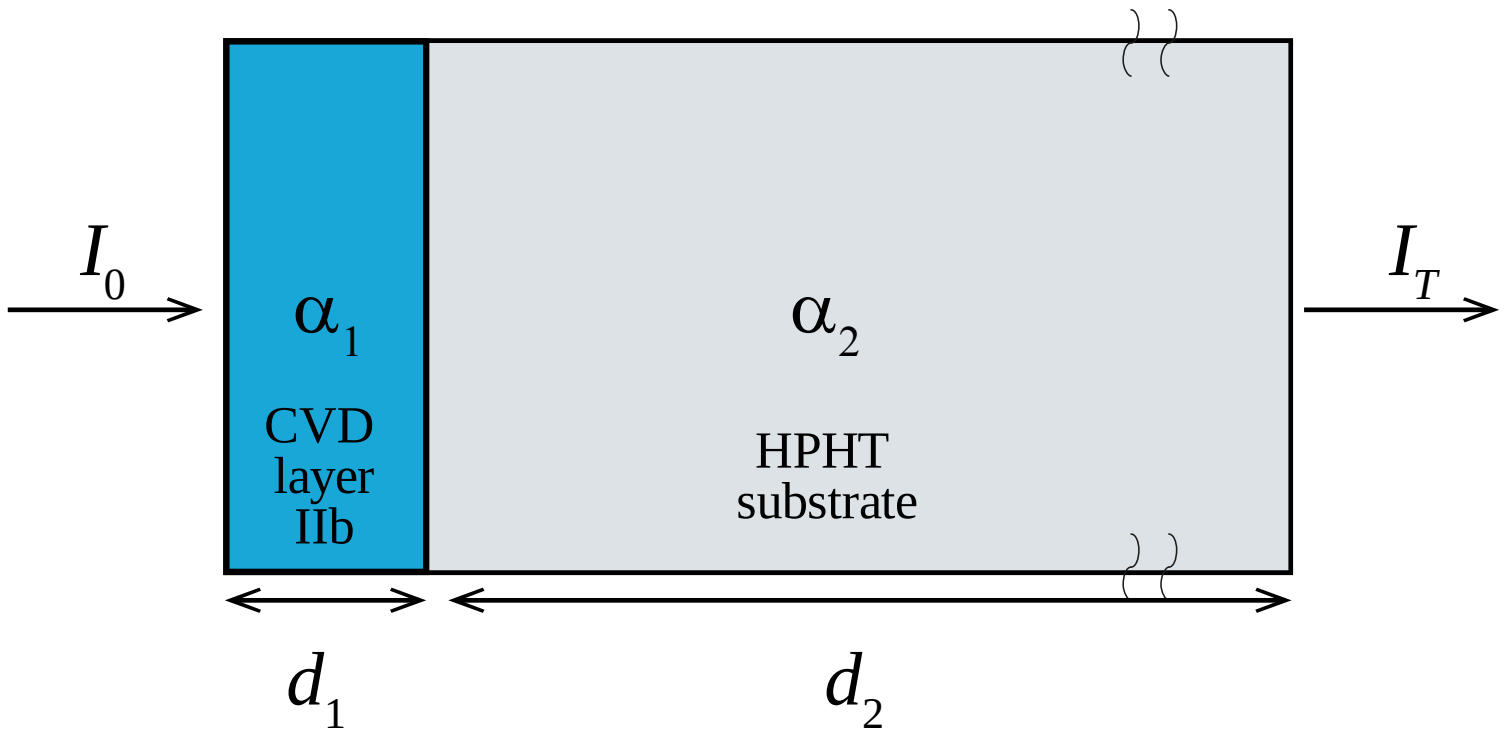

Fig. 1. Schematical representation of the diamond structure "epitaxial layer on a substrate". $I_T$ is the transmittance intensity, $T = I_T / I_0$ .

The transmittance of a multilayered structure at normal radiation incidence was calculated in [17]. Here, we use the proposed approach to derive the transmittance of the considered 2-layers structure. Assume the reflectances at the edges of the structure are equal. There is no reflection at the interface due to the same refractive coefficient of the layers. After conversion and transformation of matrix in eq. (7) [17], we obtain for the considered two-layer structure:

$$T = \frac{(1-R)^2 \exp\left[-(\alpha_1 d_1 + \alpha_2 d_2)\right]}{1 - R^2 \exp\left[-2(\alpha_1 d_1 + \alpha_2 d_2)\right]} . \qquad (3)$$

Here $\alpha_1$ is the absorption coefficient of the epitaxial layer with thickness $d_1$, $\alpha_2$ is the absorption coefficient of the substrate with thickness $d_2$.

For the typical registered spectra, the denominator in (3) is almost (with only a few percent difference) equal to one. Therefore, the denominator will be neglected further:

$$T \cong (1-R)^2 \exp\left[-(\alpha_1 d_1 + \alpha_2 d_2)\right]. \tag{4}$$

Hence

$$\alpha_1 d_1 + \alpha_2 d_2 = \ln\frac{(1-R)^2}{T} = \alpha d^*. \tag{5}$$

This expression is similar to (2), but includes the total optical density, which we will call here the *effective optical density* $\alpha d^*$. Note that in the literature, the optical density $D$ is often calculated as the decimal logarithm of the ratio of the incident to the transmitted flux, i.e. $\alpha d = D\ln 10$.

It is clear that the absorption coefficient depends on the wavelength, and for a particular wavelength the considered structure can absorb photons by boron atoms or by lattice. In addition, the substrates used for CVD growth may contain uncontrolled nitrogen impurity causing the absorption at N-specific wavelengths. Consider three cases (see, for clarity, Fig. 3):

1) Absorption in the peak near 2000 $cm^{-1}$. In this wavelength region, there is absorption by phonons of the crystal lattice in both the substrate (with $\alpha_{2\Omega}$) and the epitaxial layer (with $\alpha_{1\Omega}$). For this wavelength, we rewrite (5) as:

$$\alpha d^*_{2000} = \alpha_{1\Omega} d_1 + \alpha_{2\Omega} d_2. \tag{6}$$

There is no absorption by boron atoms: $\alpha_{1B} d_1 = \alpha_{2B} d_2 = 0$.

2) Absorption in the region 1000–1400 $cm^{-1}$. Here, the absorption by single nitrogen atoms N (C defect) and nitrogen pairs N-N (A defect) generally takes place [18]. In addition, the 1290 $cm^{-1}$ peak arises due to boron absorption in the heavily doped epitaxial layer, i.e.:

$$\alpha d^*_{1290} = \alpha_{1B} d_1 + \alpha_{1N} d_1 + \alpha_{2N} d_2. \tag{7}$$

3) Absorption in the peak near 2800 $cm^{-1}$. At this wavelength, boron atoms absorb with absorption coefficient $\alpha_{1B}$ in the epitaxial layer of thickness $d_1$. In addition, it is necessary to take into account the relatively small phonon absorption in the epitaxial layer and in the substrate $\alpha_{1\Omega}d_1 + \alpha_{2\Omega}d_2$:

$$\alpha d^{*}_{2800} = \alpha_{1B}d_1 + \alpha_{1\Omega}d_1 + \alpha_{2\Omega}d_2 . \quad (8)$$

The same consideration holds true for the 2458 $cm^{-1}$ band.

In the obtained expressions the sought-for value is $\alpha_{1B}$. The values $\alpha_{1\Omega}$ and $\alpha_{2\Omega}$ are also determined from experiments, and, as the real experience shows, they may be different. The coefficient *R* is taken from the corresponding experimental reflection spectra.

## Material and methods

We have investigated a set of single-crystal homoepitaxial CVD diamond samples and HPHT substrates. HPHT substrates were laser cut from diamond single crystals perpendicularly to [100] crystallographic direction. Crystals were grown by New Diamond Technology and NPK Almaz (Russia) and were undoped as well as intentionally boron- or nitrogen-doped. After further polishing and cleaning these substrates were used for deposition of boron-doped CVD homoepitaxial layers. The layers were grown in a plasma chemical reactor based on a cylindrical resonator under conditions with a high specific energy input, which makes it possible obtaining diamond layers of high crystalline perfection at a sufficiently high growth rate of ~4 μm/h [19]. A solution of trimethylborate $B(OCH_3)_3$ in ethanol was used for B doping. Samples CVD1, CVD5 were grown on substrates containing nitrogen in the form of A-centers (type IaA), samples CVD3, CVD4, and CVD6 were grown on substrates containing nitrogen in the form of C-centers (type Ib).

Transmission and reflection spectra of diamond structures with CVD epilayers and HPHT substrates in the near- and mid-infrared regions (700 – 4000 $cm^{-1}$) were measured using the Nikolet-6700 FTIR spectrometer with a normal reflection unit (Thermo Fisher Scientific, USA). Some measurements were performed in the range from 600 to 7000 $cm^{-1}$ using Vertex 70 FT-IR spectrometer with Hyperion 1000 IR microscope (Bruker, USA) with a resolution of 4 $cm^{-1}$ and averaging of 32 scans.

**Results**

Correctly determining the thickness of the thin epitaxial CVD layer is crucial because it is the only layer involved in the absorption of photons by boron atoms. In optics, this can be performed by analyzing the spectral interference in the experimental spectrum, arising due to the reflection from the layer boundaries [20]. The obvious approach of finding this value directly from the IR absorption spectra does not solve the problem, since the recorded interference signal is very weak. Consequently, we have additionally measured the reflection spectra of the CVD samples. The determination of the interference period is illustrated using the data for sample CVD3 in Fig. 2.

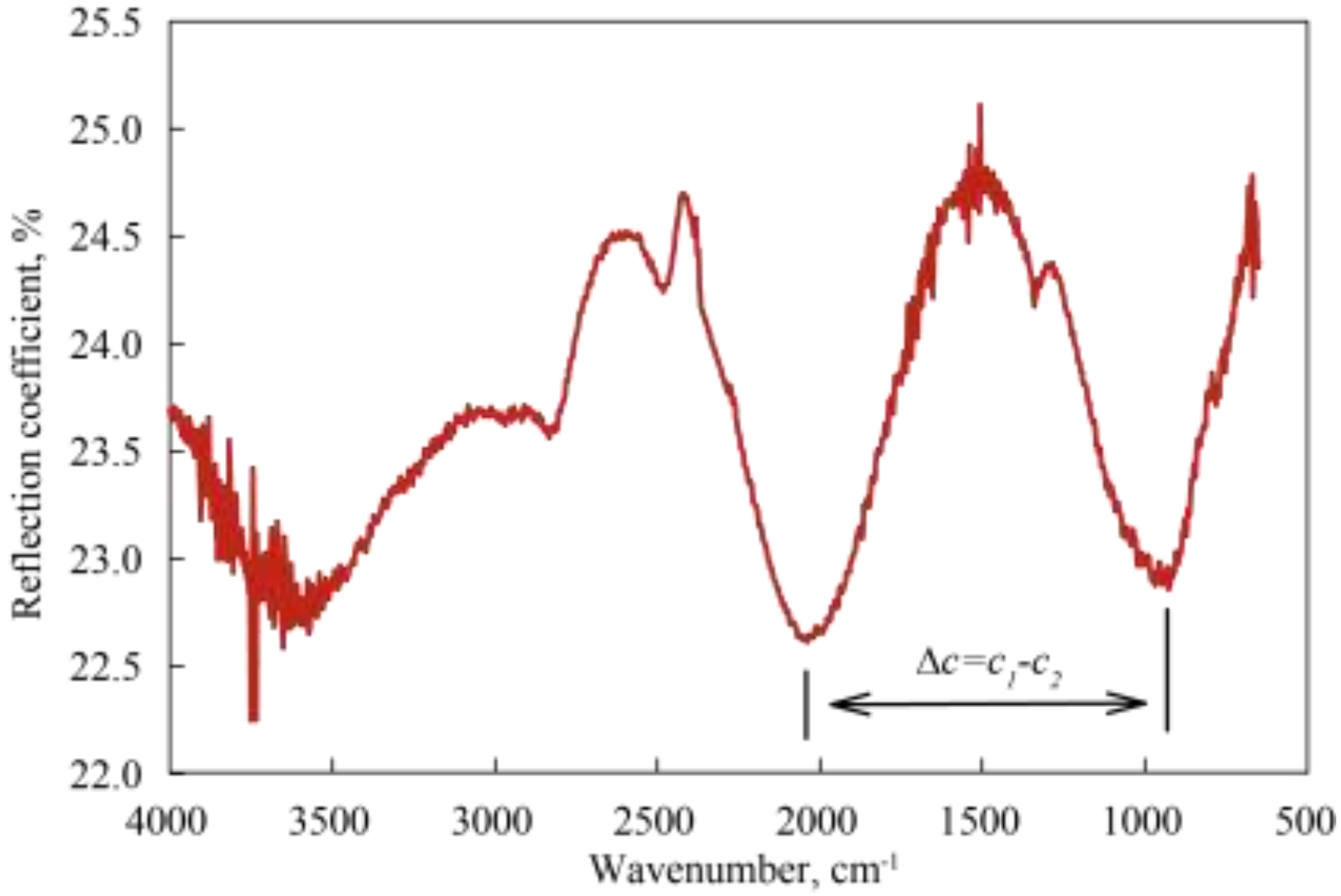

Fig. 2. Determination of the interference period for CVD diamond layer from the reflection spectra.

The thickness of the epilayer was calculated by the formula [21]:

$$d_{\mathrm{epi}} = \frac{1}{n_{\mathrm{diam}}} \cdot \frac{1}{2\Delta c}, \tag{9}$$

where $n_{\mathrm{diam}} \approx 2.4$ is the diamond refractive index [4,22,23], and $\Delta c$ is the period of interference oscillations.

The interference period is inversely proportional to the layer thickness. Therefore, for typical thicknesses of epilayers of several μm, from 2 to 4 interference periods are observed in the considered part of the reflection spectra. For comparison, the reflection spectra of bulk samples usually contain several tens of periods [24]. The epilayer thicknesses of all studied homoepitaxial samples were derived in this manner. They were 2.0 – 2.72 μm, which coincided well with the growth data.

Effective optical density spectra of the homoepitaxial diamond structures CVD1, CVD3, CVD4, and CVD5 are presented in Fig. 3. For samples CVD3 and CVD4 (grown on Ib substrates) we have determined the nitrogen concentration $N_C$ = $2.4 \cdot 10^{19}$ $cm^{-3}$ from the amplitude of the 1130 $cm^{-1}$ absorption peak using the established algorithm for the C-defect ($[N_C] = 25 \cdot H_{1130}$) from [16] ).

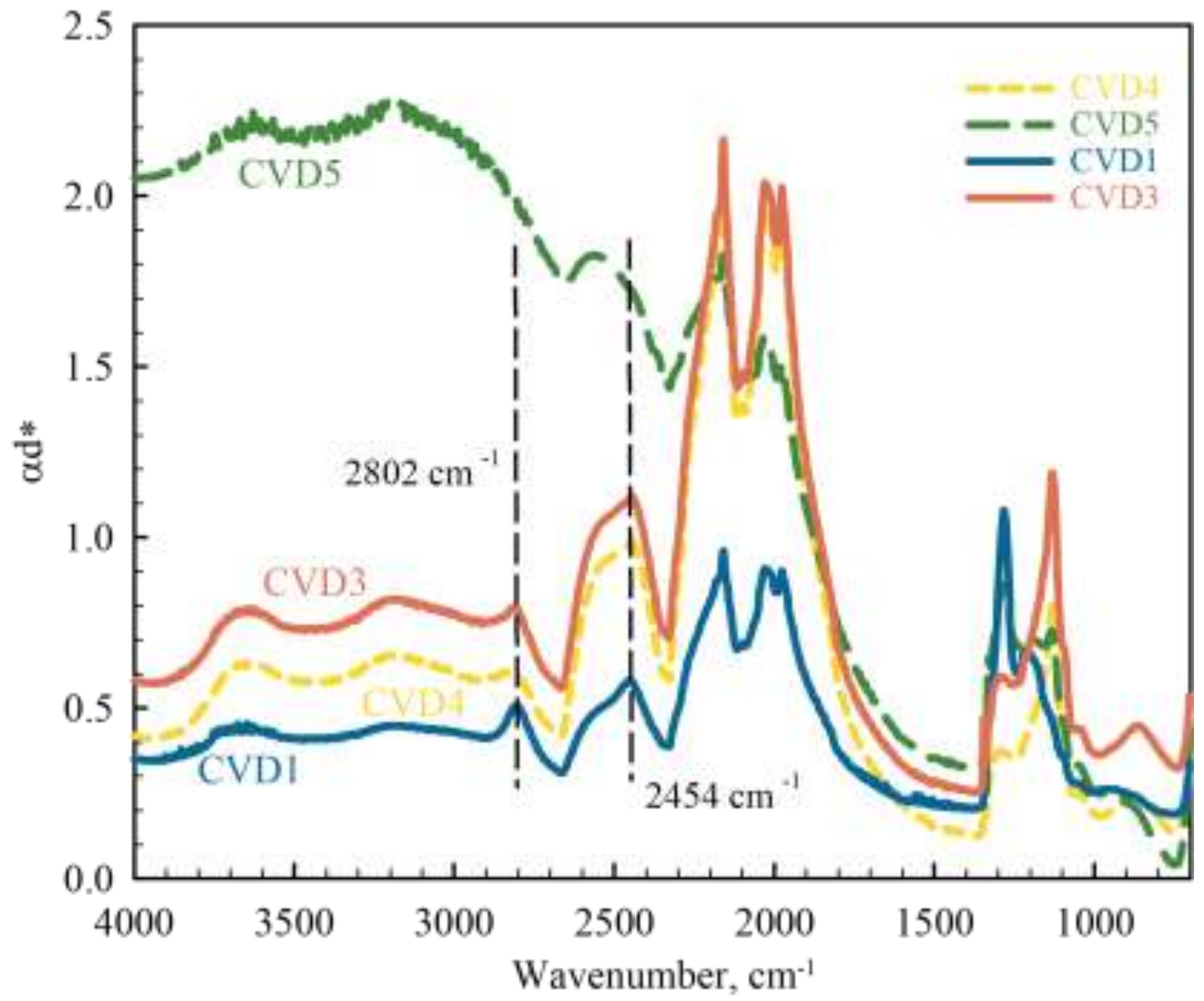

Fig. 3. FTIR optical density spectra of homoepitaxial structures CVD1, CVD3, CVD4, CVD5.

Analysis of FTIR spectra shows that in the samples CVD1, CVD3, and CVD4 the expected absorption bands of boron atoms at 2800 $cm^{-1}$ and 2454 $cm^{-1}$ are observed against a background of strong lattice absorption.

For sample CVD5 with ultrahigh boron concentration ($>5 \cdot 10^{19}$ $cm^{-3}$), the characteristic peaks in the considered regions become indistinguishable due to huge lattice absorption. The authors [11] recommend to determine boron concen-tration using the intensity of an additional (the ternary) absorption peak at 1290 $cm^{-1}$, but in our case this peak is superimposed by the absorption of residual single nitrogen atoms, and the concentration evaluation becomes difficult.

**Method for determination of boron concentration**

To calculate the boron concentration in the epitaxial CVD layer, it seems natural subtracting the optical density (or absorption) spectrum of the initial boron-free HPHT substrate from that of the whole homoepitaxial sample, and the thickness of the epilayer can be neglected here. However, as can be seen from the comparison of these spectra (Fig. 4), in addition to the registered boron absorption peaks at 2800 and 2454 $cm^{-1}$, the structure with the epitaxial layer has an increased lattice absorption in the whole considered wavenumber range, which makes this approach incorrect.

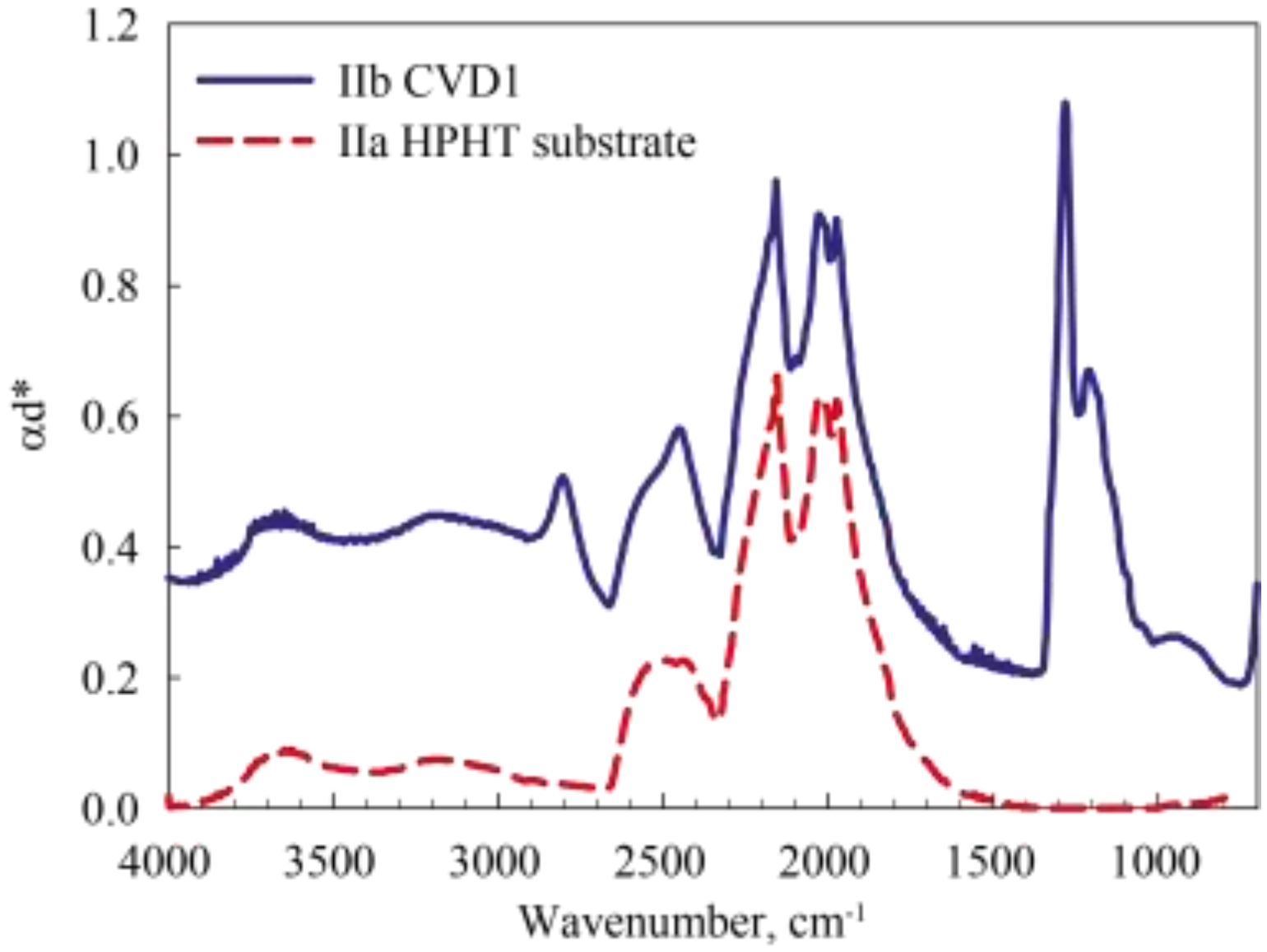

Fig. 4. FTIR optical density spectra of a structure "substrate + boron-doped epitaxial layer", sample CVD1 (solid line) compared to undoped substrate (dashed line).

Moreover, a significant modification of the homoepitaxial structure spectrum in the one-phonon absorption region 1350 – 700 $cm^{-1}$ is observed. Such absorption (theoretically impossible in a perfect diamond-like semiconductor for symmetric reasons) can arise due to the violation of local crystal symmetry, which is associated with significantly different growth conditions in CVD and HPHT processes as well with the introduction of various background impurities into the growing epitaxial layer [6,25,26].

In contrast, FTIR spectra of boron doped and non-doped bulk samples (substrates) show about the same amplitude in the two-phonon absorption region and no response in the one-phonon absorption region, Fig. 5.

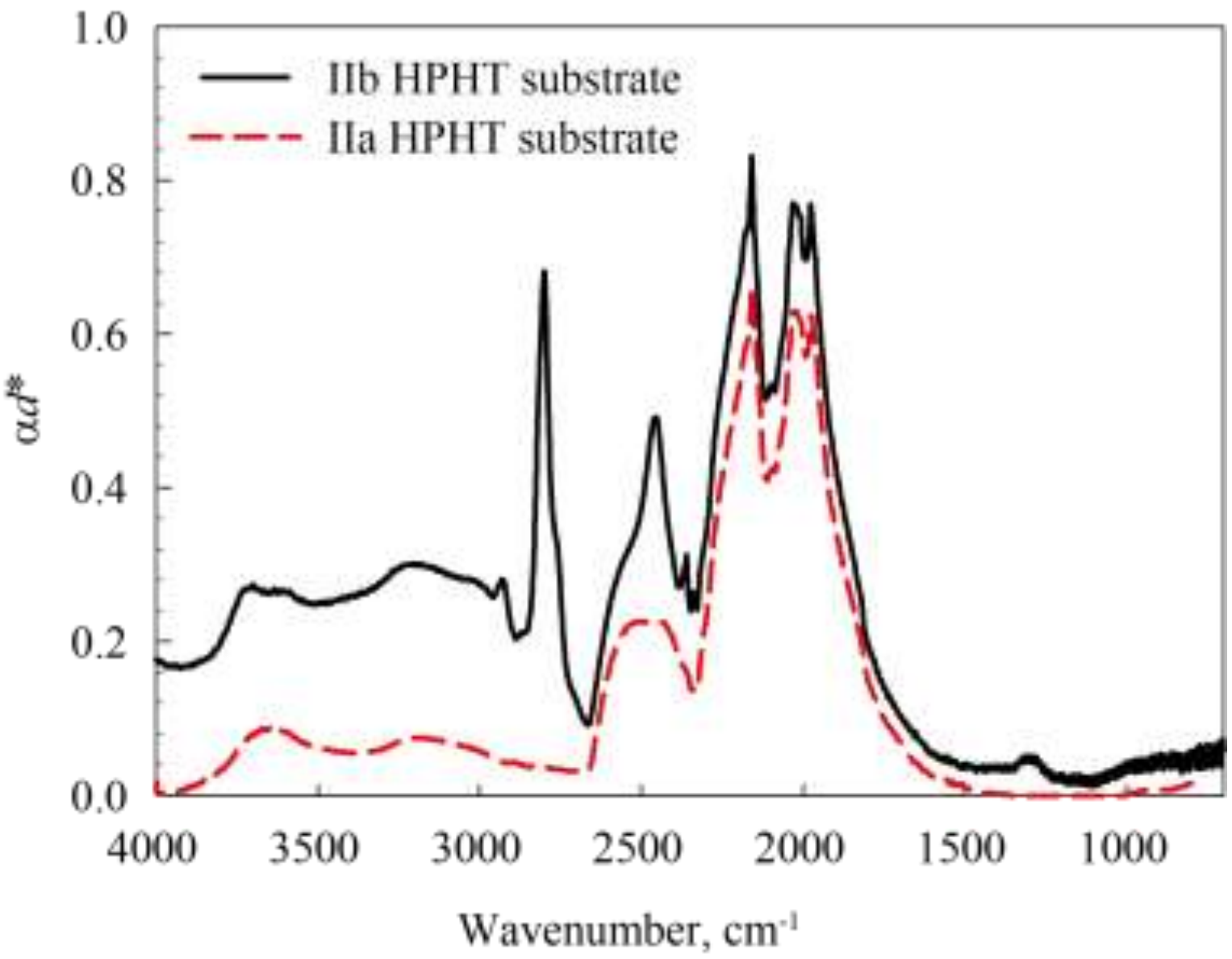

Fig. 5. FTIR optical density spectra of a weakly boron-doped IIb HPHT plate (solid line) compared to a type IIa HPHT substrate (dashed line). The thicknesses of the samples are equal.

So, further we suggest the technique for correct determination of boron concentration in homoepitaxial diamond layers based on accurate analysis of FTIR data and using the algorithm implemented in [11] for correct peak selection. The essence is as follows. At the first step, the reflection spectrum is measured and the thickness of the epitaxial layer is determined. At the second step, the IR transmission spectrum is measured in a wide range of wavelengths. The resulting spectrum is analyzed and divided into separate sections corresponding to different absorption mechanisms. Normalization by the appropriate thickness of the absorbing layer allows one to correctly calculate the absolute concentration of the impurity at the final step.

Consider the application of the algorithm for the sample CVD6 (grown on Ib substrate). Three mentioned regions can be clearly distinguished in the FTIR transmission spectrum (Fig. 6):

1) two-phonon lattice absorption (wavenumber range 1400...2350 $cm^{-1}$);

2) absorption by boron atoms (2350...2980 $cm^{-1}$);

3) absorption by nitrogen atoms (1000...1400 $cm^{-1}$).

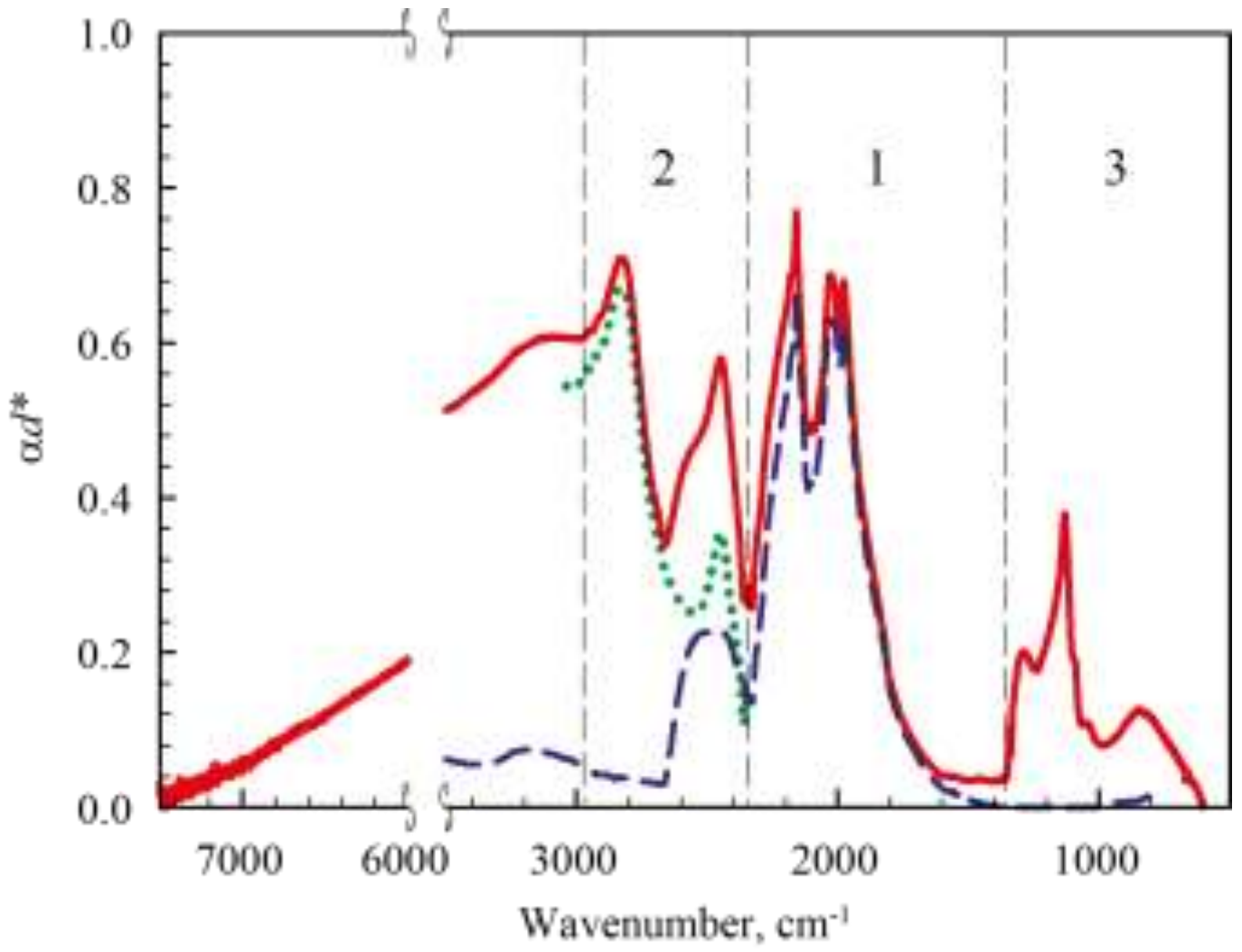

Fig. 6. Three distinctive regions of the FTIR optical density spectrum: a) CVD6 sample with boron-doped CVD layer (solid line); b) undoped HPHT substrate of the same thickness (dashed line). Dotted plot shows CVD6 spectrum with subtracted phonon absorption.

Again, in diamond crystals possessing purely covalent bonding, no single-phonon IR absorption is noticed [27] due to the absence of dipole moment [4,28].

When converting the measured effective optical density into the absorption coefficient, a different thickness *d* should be taken for each region as was discussed above. The initial substrate thickness was $d_{sub}$ = 500 µm. The estimated (by IR reflectance spectra) thickness of the epilayer was $d_{epi}$ = 2.7 µm. The derived absorption spectra for these three selected regions are shown in Fig. 7. Obviously, the large value of the absorption coefficient in case *a*) is achieved due to the small thickness of the epitaxial layer.

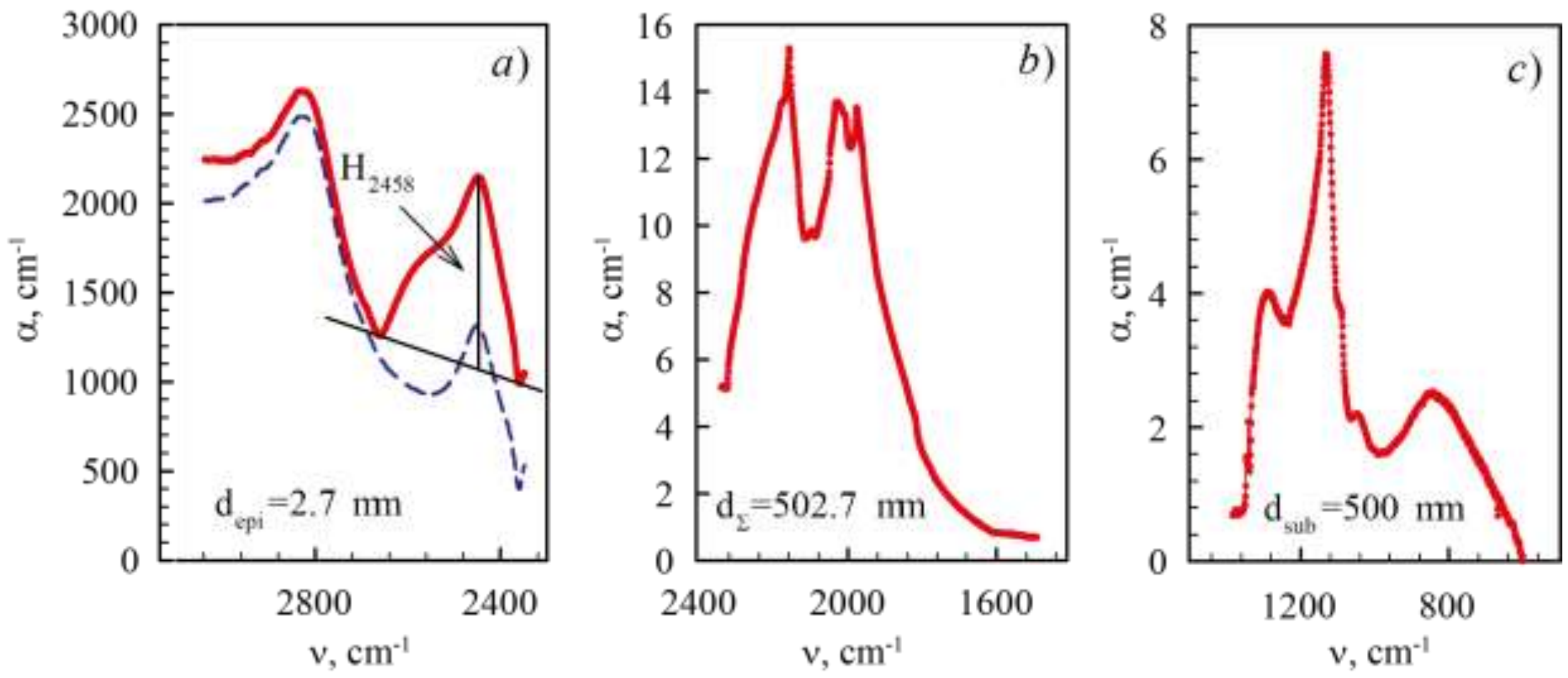

Fig. 7. Absorption spectra of CVD6 sample for different wavenumber regions: a) absorption by boron atoms (dashed line shows spectrum with subtracted phonon absorption), b) two-phonon absorption, c) absorption by single nitrogen atoms.

The boron concentration was determined using the abovementioned algorithm assuming the baseline subtraction [11], Fig.7a). The calculation gives $N_B = 1.9 \cdot 10^{19}$ cm$^{-3}$.

Note that the sections of the absorption spectrum of the undoped substrate near the characteristic boron lines at 2802 and 2452 cm$^{-1}$ have almost the same amplitudes, so the subtraction of the baseline absorption at phonons (practiced by some authors) here does not result in an increased accuracy of the concentration determination.

**Analysis and discussion**

In the same manner, we have analyzed the optical spectra of the studied CVD samples. The results obtained are presented in Table 1. It seems fairly reasonable to calculate the boron concentration from the amplitude of the boron absorption peak using the calibration coefficients (1) approved for bulk crystals, since it is confirmed by the careful and laborious calibration performed by various authors, in particular [2,11,16].

Table 1. Growth and estimated characteristics of the studied homoepitaxial diamond structures

| Characteristics | Sample No. | | | | | |
|---|---|---|---|---|---|---|
| | CVD1 | CVD2 | CVD3 | CVD4 | CVD5 | CVD6 |
| B/C ratio in gas phase, ppm | 600 | 600 | 2300 | 9200 | 18000 | 9500 |
| O/C ratio in gas phase, ppm | 0.16 | 0.04 | 0.04 | 0.16 | 0.16 | 0.5 |
| Substrate thickness, mm | 0.5 | 1.5 | 1 | 1.5 | 0.5 | 0.5 |
| Epilayer thickness (*d*), μm | 2.71 | 2.29 | 2.06 | 2.11 | 2.30 | 2.7 |
| Boron concentration [B], $cm^{-3}$ | $2.3 \cdot 10^{18}$ | | $4 \cdot 10^{18}$ | $2.9 \cdot 10^{18}$ | n/a | $1.9 \cdot 10^{19}$ |
| Hole concentration (*p*), $cm^{-3}$ | $4 \cdot 10^{18}$ | $3 \cdot 10^{18}$ | $6.2 \cdot 10^{18}$ | $6.5 \cdot 10^{18}$ | $4 \cdot 10^{19}$ | $1.6 \cdot 10^{19}$ |

Further, we have compared the optically estimated concentration with the measurements performed by capacitance-voltage (C-V) profiling, which is widely and successfully utilized for similar purposes. For ultrawide bandgap semiconductors like diamond, the registered by C-V technique concentration fundamentally depends on the probing frequency, since boron is weakly ionized. This peculiarity, known as the frequency dispersion of the capacitance-voltage characteristics of wide bandgap semiconductors, provides rather the "apparent" free charge carrier concentration, not merely the doping level [29–32]. To take this feature into account, we have performed the measurements at the lowest available frequencies (typically 10-50 kHz), where the dispersion is weakly pronounced, and the apparent charge carrier concentration almost approaches the impurity concentration.

For C-V measurements, Ohmic and rectifying contacts were fabricated on the top of epitaxial layers. The measurements were performed using the automated admittance spectroscopy setup based on the Agilent 4980A RLC meter and Janis cryogenic probe station [33]. Note two differences between optical FTIR and electrical C-V measurements. Firstly, the C-V technique does provide measurements of boron concentration in highly doped epitaxial layers, when the primary and sec-ondary FTIR peaks of boron are not observed (sample CVD5, see also Fig. 3). Secondly, C-V characteristics provide a spatial distribution (within about upper 100 nm of the layer) of the concentration, while optical measurements provide only integral concentration. The *p* values in Table 1 are the apparent C-V concentration profiles averaged in depth. In general, these values have adequate coincidence with the measured boron concentrations within the designated features and difficulties of the methods.

## Conclusion

The paper critically analyzes the existent approaches for determination of boron impurity concentration by optical methods, primarily by the FTIR technique. The historically established classification of diamonds into types I and II according to the transparency of samples to UV radiation covers the concentration interval of approximately $5\cdot10^{16}$ –$1\cdot10^{18}$ cm$^{-3}$. This range is much narrower than the boron concentration variety suitable for application in modern microelectronic devices.

In relation to the intensive development of the epitaxial CVD growth of single-crystal diamond layers on homogeneous substrate, we have extended the technique for determination the boron concentration by nondestructive FTIR spectra for doped epitaxial layers. The peculiarities of the considered structures are the passage of probing radiation through a two-layer structure with different thicknesses of the absorbing media. For this situation, an expression for the effective optical density is obtained. The correct amplitudes of the boron absorption peaks were evaluated for a set CVD homoepitaxial layers grown on HPHT substrate. To calculate the boron concentration from the derived amplitude we used the calibration coefficients applied for bulk doped samples.

The experimental FTIR results were thoroughly compared with precision measurements of concentration by the capacitance-voltage method, stating their adequate coincidence.

## Acknowledgements

The authors give thanks to Prof. A. L. Vikrarev, Dr. S. Bogdanov and their colleagues (Institute of Applied Physics of the Russian Academy of Sciences) for the provided CVD diamond samples.

**References**

[1] C.D. Clark, R.W. Ditchburn, H.B. Dyer, The absorption spectra of natural and irradiated diamonds, Proc. R. Soc. London. Ser. A. Math. Phys. Sci. 234 (1956) 363–381. https://doi.org/10.1098/rspa.1956.0040.

[2] A.T. Collins, A.W.S. Williams, The nature of the acceptor centre in semiconducting diamond, J. Phys. C Solid State Phys. 4 (1971) 1789. https://doi.org/10.1088/0022-3719/4/13/030.

[3] J. Walker, Optical absorption and luminescence in diamond, Reports Prog. Phys. 42 (1979) 1605–1659. https://doi.org/10.1088/0034-4885/42/10/001.

[4] M.E. Thomas, W.J. Tropf, Optical properties of diamond, Johns Hopkins APL Tech. Dig. 14 (1993) 16–23. https://doi.org/10.1201/9780429283260-7.

[5] T. Petit, L. Puskar, FTIR spectroscopy of nanodiamonds: Methods and interpretation, Diam. Relat. Mater. 89 (2018) 52–66. https://doi.org/10.1016/j.diamond.2018.08.005.

[6] R.P. Mildren, J.R. Rabeau, Optical Engineering of Diamond, Wiley- VCH Verlag GmbH & Co. KGaA, 2013. https://doi.org/10.1002/9783527648603.

[7] A.M. Zaitsev, Optical properties of diamond: a data handbook, Springer Berlin, Heidelberg, 2001. https://doi.org/https://doi.org/10.1007/978-3-662-04548-0.

[8] S. Koizumi, H. Umezana, J. Pernot, M. (Eds) Suzuki, Power Electronics Device Applications of Diamond Semiconductors, Elsevier, 2018. https://doi.org/10.1016/c2016-0-03999-2.

[9] R.M. Chrenko, Boron, the dominant acceptor in semiconducting diamond, Phys. Rev. B. 7 (1973) 4560–4567. https://doi.org/10.1103/PhysRevB.7.4560.

[10] A.T. Collins, Colour centres in diamond, J. Gemmol. 18 (1982) 37–75.

https://doi.org/10.15506/JOG.1982.18.1.37.

[11] D. Howell, A.T. Collins, L.C. Loudin, P.L. Diggle, U.F.S. D'Haenens-Johansson, K. V. Smit, A.N. Katrusha, J.E. Butler, F. Nestola, Automated FTIR mapping of boron distribution in diamond, Diam. Relat. Mater. 96 (2019) 207–215. https://doi.org/10.1016/j.diamond.2019.02.029.

[12] E. Gaillou, J.E. Post, D. Rost, J.E. Butler, Boron in natural type IIb blue diamonds: Chemical and spectroscopic measurements, Am. Mineral. 97 (2012) 1–18. https://doi.org/10.2138/am.2012.3925.

[13] J.M. King, T.M. Moses, J.E. Shigley, C.M. Welbourn, S.C. Lawson, M. Cooper, Characterizing natural-color type IIb blue diamonds, Gems Gemol. 34 (1998) 246–268. https://doi.org/10.5741/GEMS.34.4.246.

[14] T.H. Borst, O. Weis, Boron-doped homoepitaxial diamond layers: Fabrication, characterization, and electronic applications, Phys. Status Solidi Appl. Res. 154 (1996) 423–444. https://doi.org/10.1002/pssa.2211540130.

[15] V.S. Bormashov, S.A. Tarelkin, S.G. Buga, M.S. Kuznetsov, S.A. Terentiev, A.N. Semenov, V.D. Blank, Electrical properties of the high quality boron-doped synthetic single-crystal diamonds grown by the temperature gradient method, Diam. Relat. Mater. 35 (2013) 19–23. https://doi.org/10.1016/j.diamond.2013.02.011.

[16] D. Howell, C.J. O'Neill, K.J. Grant, W.L. Griffin, N.J. Pearson, S.Y. O'Reilly, μ-FTIR mapping: Distribution of impurities in different types of diamond growth, Diam. Relat. Mater. 29 (2012) 29–36. https://doi.org/10.1016/j.diamond.2012.06.003.

[17] S.H. Wemple, J.A. Seman, Optical Transmission Through Multilayered Structures, Appl. Opt. 12 (1973) 2947. https://doi.org/10.1364/ao.12.002947.

[18] C. Breeding, J. Shigley, The "type" classification system of diamonds and its importance in gemology, Gems Gemol. 45 (2009) 96–111. https://doi.org/10.5741/GEMS.45.2.96.

[19] S.A. Bogdanov, A.L. Vikharev, M.N. Drozdov, D.B. Radishev, Synthesis of thick and high-quality homoepitaxial diamond with high boron doping level:

Oxygen effect, Diam. Relat. Mater. 74 (2017) 59–64. https://doi.org/10.1016/j.diamond.2017.02.004.

[20] M.F. Panov, V.P. Rastegaev, S.A. Korlyakova, Spectral interference in a carbide-silicon n--n + structure, Tech. Phys. 59 (2014) 1252–1254. https://doi.org/10.1134/S1063784214080179.

[21] D.K. Schroder, Semiconductor Material and Device Characterization: Third Edition, John Wiley & Sons, Inc., 2005. https://doi.org/10.1002/0471749095.

[22] E.D. Palik, Handbook of optical constants of solids, 2012. https://doi.org/10.1016/C2009-0-20920-2.

[23] G. Turri, S. Webster, Y. Chen, B. Wickham, A. Bennett, M. Bass, Index of refraction from the near-ultraviolet to the near-infrared from a single crystal microwave-assisted CVD diamond, Opt. Mater. Express. 7 (2017) 855. https://doi.org/10.1364/ome.7.000855.

[24] Afanasjev A. V., Zubkov V. I., Ilyin V. A., Luchinin V. V., Pavlova M. V., Panov M. F., Trushliakova V. V., Firsov D. D., Determination of thickness and doping features of multilayer 4H-SiC structures by frequency analysis of IR reflection spectra, Tech. Phys. Lett. 48 (2022) 74. https://doi.org/10.21883/tpl.2022.01.52476.19012.

[25] R. Issaoui, A. Tallaire, A. Mrad, L. William, F. Bénédic, M.A. Pinault-Thaury, J. Achard, Defect and Threading Dislocations in Single Crystal Diamond: A Focus on Boron and Nitrogen Codoping, Phys. Status Solidi Appl. Mater. Sci. 216 (2019). https://doi.org/10.1002/pssa.201900581.

[26] V. Ţucureanu, A. Matei, A.M. Avram, FTIR Spectroscopy for Carbon Family Study, Crit. Rev. Anal. Chem. 46 (2016) 502–520. https://doi.org/10.1080/10408347.2016.1157013.

[27] A.T. Collins, Intrinsic and extrinsic absorption and luminescence in diamond, 185 (1993) 284–296.

[28] P.Y. Yu, M. Cardona, Fundamentals of Semiconductors, Springer Berlin, Heidelberg, 1996. https://doi.org/10.1007/b137661.

[29] D.S. Frolov, V.I. Zubkov, Frequency dispersion of capacitance-voltage

characteristics in wide bandgap semiconductor-electrolyte junctions, Semicond. Sci. Technol. 31 (2016) 125013. https://doi.org/10.1088/0268-1242/31/12/125013.

[30] A. Solomnikova, V. Lukashkin, V. Zubkov, A. Kuznetsov, A. Solomonov, Carrier concentration variety over multisectoral boron-doped HPHT diamond, Semicond. Sci. Technol. 35 (2020). https://doi.org/10.1088/1361-6641/ab9a5f.

[31] V.I. Zubkov, A.V. Solomnikova, A.V. Solomonov, A.V. Koliadin, J.E. Butler, Characterization of boron-doped single-crystal diamond by electrophysical methods (review), Tech. Phys. 68 (2023) 3–25.

[32] G.H. Glover, The C-V characteristics of Schottky barriers on laboratory grown semiconducting diamonds, Solid State Electron. 16 (1973) 973–978. https://doi.org/10.1016/0038-1101(73)90196-2.

[33] V.I. Zubkov, O. V. Kucherova, S.A. Bogdanov, A. V. Zubkova, J.E. Butler, V.A. Ilyin, A. V. Afanas’Ev, A.L. Vikharev, Temperature admittance spectroscopy of boron doped chemical vapor deposition diamond, J. Appl. Phys. 118 (2015). https://doi.org/10.1063/1.4932664.